\documentclass[prd,nofootinbib,onecolumn,superscriptaddress]{revtex4}

\usepackage{amsmath, amssymb, amsthm, graphicx, epsfig, fancyhdr,epsfig, slashed, mathrsfs}

\usepackage{natbib}
\usepackage{float}

\usepackage{mathtools} 

\usepackage{tikz,xcolor,hyperref}

\definecolor{lime}{HTML}{A6CE39}
\DeclareRobustCommand{\orcidicon}{
	\begin{tikzpicture}
	\draw[lime, fill=lime] (0,0) 
	circle [radius=0.2] 
	node[white] {{\fontfamily{qag}\selectfont \tiny ID}};
	\draw[white, fill=white] (-0.0625,0.095) 
	circle [radius=0.007];
	\end{tikzpicture}
	\hspace{-2mm}
}

\foreach \x in {A, ..., Z}{\expandafter\xdef\csname orcid\x\endcsname{\noexpand\href{https://orcid.org/\csname orcidauthor\x\endcsname}
			{\noexpand\orcidicon}}
}

\newcommand{\be}{\begin{equation}}
\newcommand{\ee}{\end{equation}}
\newcommand{\bea}{\begin{eqnarray}}
\newcommand{\eea}{\end{eqnarray}}

\begin{document}


\title{Fate of False Vacuum in Non-perturbative Regimes}

\author{Marco Frasca\orcidA{}}
\email{marcofrasca@mclink.it (Corresponding Author}
\affiliation{Rome, Italy}

\author{Anish Ghoshal\orcidB{}}
\email{anish.ghoshal@fuw.edu.pl}

\affiliation{Institute of Theoretical Physics, Faculty of Physics, University of Warsaw,ul.  Pasteura 5, 02-093 Warsaw, Poland}

\author{Nobuchika Okada\orcidC{}}
\email{okadan@ua.edu}
\affiliation{Department of Physics and Astronomy, \\ University of Alabama, Tuscaloosa, AL 35487, USA}

\begin{abstract}
\textit{We use some exact results in the scalar field theory to revise the analysis by Coleman and Callan about the false vacuum decay and propose a simple non-perturbative formalism. We introduce exact Green's function which incorporates non-perturbative corrections in the strong coupling regimes of the theory. The solution of the scalar field theory involves Jacobi elliptical function and has been used to calculate the effective potential for any arbitrary coupling values. We demonstrate the use of this formalism in a simple $\lambda \phi^4$ theory and show that the effective potential exhibits a false minimum at the origin. We then calculate the false vacuum decay rate, in the thin wall approximation, and suggest simple analytic formulas which may be useful for the analysis for the first order phase transition beyond the perturbative regime. In our methodology, we show that the standard results obtained in perturbation theory are reproduced by taking the coupling values very small.}
\end{abstract}

\maketitle


\section{Introduction}

The nature of Higgs false vacuum and its stability has been a topic of great interest since the 1970s with seminal papers by Coleman \cite{Coleman:1977py,Callan:1977pt}. In the last decade, the arrival of gravitational wave astronomy and the discovery of Higgs metastability after the discovery of the Higgs boson at the LHC have led to increased interests \cite{Degrassi:2012ry,Isidori:2001bm}. LIGO has put bounds \cite{2009Natur.460..990A} and very soon we should also be able to look
at the stochastic gravitational wave background (SGWB) produced from primordial first-order phase transition, so that accurate calculations of the false vacuum (FV) decay rate in myriad of beyond the Standard Model (BSM) theories have become of utmost importance.  

Currently, the established formalism, developed by Callan and Coleman \cite{Coleman:1977py,Callan:1977pt}, uses the saddle point approximation in order to write the imaginary part of the false vacuum energy in terms of bounce solutions of the tree-level action. Extension to loop-level bounce calculations \cite{Weinberg:1992ds}, both with perturbative and non-perturbative approaches to incorporate radiative corrections, have been mostly limited to weakly-coupled theories. So, admitting an actual bounce beyond tree-level has been extremely challenging where it does not exist at the tree-level.
For strongly-coupled theories we are still in quest to find out new methods for false vacuum decay which will be accurate, versatile, and robust, and may lead to applicable results at the calculation of phase transitions and generation of Gravitational Waves. Such an attempt was made recently in Ref. \cite{Croon:2021vtc}; their method relied on functional renormalization group (FRG) techniques developed in Ref. \cite{Wetterich:1992yh} via inventing quasi-stationary effective action. The authors argued that 
the proper generalization of the saddle-point method already defines a \textit{quasi-stationary effective action} (QSEA) in the form of course-grained action which only takes into account fluctuations around the background smaller than its characteristic size \footnote{For other works on bubble nucleation, etc. see Refs.\cite{Strumia:1998vd} and \cite{Plascencia:2015pga}.}.

In this paper, we revert to utilising exact solutions to the Higgs theory, found in terms of Jacobi elliptical functions, following Ref.\cite{Frasca:2015yva}. Since the Green's functions of the theory are analytically expressed, we can compute the effective action via the partition function of the theory and subsequently the false vacuum decay rate in terms of Jacobi elliptical function. The result obtained in this way is valid even in the strongly coupled regimes. For recent applications of such a technique to QCD and confinement and hadronic contributions to muon magnetic moment and dark energy cosmology, see Refs.\cite{Frasca:2021yuu,Frasca:2021mhi,Frasca:2022lwp,Chaichian:2018cyv,Frasca:2022vvp} and to the Higgs sector of the Standard Model see Ref.\cite{Frasca:2015wva}. Our scheme is also applicable to the first-order phase transition phenomena in the strongly coupled condensed matter systems.
We emphasize that this exact solution we work with for the scalar field theory is not unique for reasons that we will explain later in the text. Anyway,  a solution in closed analytical form is helpful to better understand physics and can make clearer how some results, obtained only through lattice computations so far, could be better interpreted. For this reason, we consider it worthwhile to present the analysis for the false vacuum decay for a scalar field where such an analytical solution is available.
%

This paper is organised as follows: first we discuss the formalism of calculating effective potential by using the exact solution and partition function of the theory. Next we review the standard method to compute bounce solution. After that, we introduce the exact solution in terms of Green's function via effective potential approach and then we show the results of our analysis and the validity of the results in the non-perturbative regimes. Finally we conclude with our investigation and some possible application of the results obtained.

\medskip

\section{Dyson-Schwinger equations for the scalar field}
\label{Dyson}


In this section we present material taken from a rather recent proposed solution for the hierarchy
of Dyson-Schwinger equations both for the scalar field and Yang-Mills theory \cite{Frasca:2015yva,Frasca:2021yuu,Frasca:2021mhi,Frasca:2022lwp,Chaichian:2018cyv}. This represents
a particular solution and, as such, it proved to yield a very precise account of the
spectrum of Yang-Mills theory as measured on lattice \cite{Frasca:2017slg} and a determination
of its beta function \cite{Chaichian:2018cyv}. The technique relies on a proposal due to
Bender, Milton and Savage in the framework of the PT-invariant theories \cite{Bender:1999ek}.
We would like to point out that the exact solution for the 1P-correlation function is very similar to a Fubini instanton \cite{Fubini:1976jm} representing a non-trivial vacuum for the theory breaking the translation invariance. Such a symmetry is indeed fully recovered in the 2P-correlation function that enters into the determination of the observables as expected from the LSZ-theorem. In the appendix we show that such a breaking of symmetry entails a zero mode for the scalar theory as it should.

We start from the following partition function for a scalar field
\begin{equation}
    Z[j]=\int[D\phi]e^{iS(\phi)+i\int d^4xj(x)\phi(x)}.
\end{equation}
The equation for the 1P-function is given by
\be
\left\langle\frac{\delta S}{\delta\phi(x)}\right\rangle=-j(x)
\ee
where
\be
\left\langle\ldots\right\rangle=\frac{\int[D\phi]\ldots e^{iS(\phi)+i\int d^4xj(x)\phi(x)}}{\int[D\phi]e^{iS(\phi)+i\int d^4xj(x)\phi(x)}}.
\ee
One has
\be
G_n^{(j)}(x_1,x_2,\ldots,x_n)=(-i)^n\frac{1}{Z}\frac{\delta^n Z}{\delta j(x_1)\delta j(x_2)\ldots\delta j(x_n)},
\ee
starting with
\be
G_1^{(j)}(x)=\frac{\langle\phi(x)\rangle}{Z[j]}.
\ee

We apply all this to a $\phi^4$ theory and get
\be
S=\int d^4x\left[\frac{1}{2}(\partial\phi)^2-\frac{\lambda}{4}\phi^4\right],
\ee
so that,
\be
\label{eq:G_1}
\partial^2\langle\phi\rangle+\lambda\langle\phi^3(x)\rangle = -j(x).
\ee
For the 1P-function is
\be
Z[j]\partial^2G_1^{(j)}(x)+\lambda\langle\phi^3(x)\rangle = -j(x),
\ee
and by definition
\be
Z[j]G_1^{(j)}(x)=\langle\phi(x)\rangle.
\ee
After derivation with respect to $j(x)$, we get
\be
Z[j][G_1^{(j)}(x)]^2+Z[j]G_2^{(j)}(x,x)=\langle\phi^2(x)\rangle,
\ee
and after another derivation step it one has:
\begin{widetext}
\be
Z[j][G_1^{(j)}(x)]^3+3Z[j]G_1^{(j)}(x)G_2(x,x)+Z[j]G_3^{(j)}(x,x,x)=\langle\phi^3(x)\rangle.
\ee
\end{widetext}
We insert this into Eqn.(\ref{eq:G_1}) to obtain
\begin{widetext}
\be
\label{eq:G1_j}
\partial^2G_1^{(j)}(x)+\lambda[G_1^{(j)}(x)]^3+3\lambda G_2^{(j)}(x,x)G_1^{(j)}(x)+G_3^{(j)}(x,x,x)=-Z^{-1}[j]j(x)
\ee
\end{widetext}
Setting $j=0$, we get the first Dyson-Schwinger equation
\be
\partial^2G_1(x)+\lambda[G_1(x)]^3+3\lambda G_2(x,x)G_1(x)+G_3(x,x,x)=0,
\ee
where we realize that quantum corrections induced a mass term as we will show that $G_2(x,x)=G_2(0)$ i.e. a constant.

Our next step is to derive Eqn.(\ref{eq:G1_j}) again with respect to $j(y)$. This gives the result
\begin{widetext}
\bea
&&\partial^2G_2^{(j)}(x,y)+3\lambda[G_1^{(j)}(x)]^2G_2^{(j)}(x,y)+
3\lambda G_3^{(j)}(x,x,y)G_1^{(j)}(x)
+3\lambda G_2^{(j)}(0)G_2^{(j)}(x,y) \nonumber \\
&&+G_4^{(j)}(x,x,x,y)=
-iZ^{-1}[j]\delta^4(x-y)-ij(x)\frac{\delta}{\delta j(y)}(Z^{-1}[j]).
\eea
\end{widetext}
By setting $j=0$, one has for the 2P-function
\bea
&&\partial^2G_2(x,y)+3\lambda[G_1(x)]^2G_2(x,y)+ \\
&&3\lambda G_3(x,x,y)G_1(x)
+3\lambda G_2(0)G_2(x,y)
+G_4(x,x,x,y)=
-i\delta^4(x-y).  \nonumber
\eea
In principle, one can iterate such a procedure to whatever desired order yielding the full hierarchy of Dyson-Schwinger equations into PDE form.

As it should be expected, low-order Dyson-Schwinger equations depend on higher-order nP-correlation functions evaluated at some value. We can fix these values properly notwithstanding these higher-order functions are not zero as we will see. This procedure can be easily justified {\it a posteriori}. Such values are e.g. $G_3(x,x,x)$, while $G_3(x,y,z,)\ne 0$
and similarly for the 4P-correlation function and higher. We will show how this work for the 3P-correlation function below, confirming the correctness of our {\it ansatz}. This condition fixes the boundary conditions unequivocally when a solution is searched for higher-order nP-correlation functions.
%
This is a relevant point in our approach and must be consistent when all the hierarchy of Dyson-Schwinger equations is properly solved. Anyway, this choice is not unique and a different one is also possible, granting a solution. We are motivated by the simplicity that is granted in this case that we know provides a closed form solution for certain. Besides, the mass spectrum obtained in this way appears practically the same as for a Yang-Mills theory, when these two theories are properly mapped, and is in close agreement with lattice results \cite{Frasca:2017slg}. In such a case we obtain an error below 1\% for different number of colors.

For the 1P-function, we observe that the one has $G_3$ evaluated only at $x$ and we set, to be verified {\it a posteriori}, $G_3(x,x,x)=0$. It can be checked by direct substitution. So, we have the exact solution, to be checked by direct substitution,
\begin{equation}\label{solG1}
G_1(x)=\pm\sqrt{\frac{2(p^2-m^2)}\lambda}\operatorname{sn}\left(p\cdot x+\theta,\kappa\right)
\end{equation}
where
\begin{equation}
p^2=\frac12\left(\sqrt{m^4+2\lambda\mu^4}+m^2\right),\qquad
\kappa=\frac{m^2-p^2}{p^2}
\end{equation}
where $\mu$ and $\theta$ are integration constants. $\operatorname{sn}(\zeta|\kappa)$ is Jacobi's elliptic function of the first kind. We have set
\be
m^2=3\lambda G_2(0)
\ee
for the gap equation arising from purely quantum effects. This will need renormalization as it takes the form
\be
\label{eq:dm}
m^2=3\lambda\int\frac{d^4p}{(2\pi)^4}G_2(p).
\ee
We emphasize that the equation we started from for the 1P-correlation function admits two solution depending on the chosen sign. This is an effect of the $Z_2$ symmetry. Both the choices are acceptable. In the following we will select the one with plus sign for our convenience but a different choice cannot change the physical conclusions both the solutions being equivalent.
This is another aspect of our approach that is not unique. We may start from a different solution for $G_1$ such as that of a soliton, and then go onto solve the hierarchy of Dyson-Schwinger equations. So, we do not claim that our choice is the one that will be found in nature. Anyway, it grants a complete solution to the Dyson-Schwinger set of equations through Jacobi elliptical functions when the other condition on higher nP-correlation functions we cited above is applied.
%

A relevant aspect of our solution is that this is similar to a Fubini instanton \cite{Fubini:1976jm} that breaks translation invariance. Anyway, as we should see in a moment, the $G_2$ correlation function preserves translation invariance granting the correct application of the LSZ theorem by scattering states to determine observables. So, we expect no measurable physical effect from this symmetry breaking. Besides, this symmetry breaking entails a zero mode for the theory as shown in the appendix.

The 2P-function is given by, in momentum space,
\begin{widetext}
\be
G_2(p)=\frac{\pi^3}{2(1-\kappa)K^3(\kappa)\sqrt{\kappa}}
\sum_{n=0}^\infty(2n+1)^2(-1)^n\frac{q^{n+\frac{1}{2}}}{1-q^{2n+1}}\frac{i}{p^2-m_n^2+i\epsilon}
\ee
\end{widetext}
where
\be
\kappa=\frac{m^2-\sqrt{m^4+2\lambda\mu^4}}{m^2+\sqrt{m^4+2\lambda\mu^4}},
\ee
and
\be
q=e^{-\pi\frac{K^*(\kappa)}{K(\kappa)}}
\ee
where $K^*(\kappa)=K(1-\kappa)$. The spectrum is given by
\be
m_n=(2n+1)\frac{\pi}{2K(\kappa)}\frac{1}{\sqrt{2}}\sqrt{m^2+\sqrt{m^4+2\lambda\mu^4}}.
\ee
We notice that we have recovered translation invariance for propagating degrees of freedom and so, $G_2(x,x)=G_2(0)$ is a constant as stated above. Then, assuming the shift $m^2$ negligibly small, these equations simplify to
\be
G_2(p)=\frac{\pi^3}{4K^3(-1)}
\sum_{n=0}^\infty(2n+1)^2\frac{e^{\left(n+\frac{1}{2}\right)\pi}}{1+e^{-(2n+1)\pi}}\frac{i}{p^2-m_n^2+i\epsilon}
\ee
but now
\be
m_n=(2n+1)\frac{\pi}{2K(-1)}\left(\frac{\lambda}{2}\right)^\frac{1}{4}\mu.
\ee
Higher-order nP-correlation functions can be obtained in a similar way. For the 3P- and 4P-functions, they have the form \cite{Frasca:2013tma}
\begin{widetext}
\begin{equation}
\label{eq:G_3}
   G_3(x,y,z)=-6\lambda\int dx_1 G_2(x-x_1)G_1(x_1)G_2(x_1-y)G_2(x_1-z),
\end{equation}
\end{widetext}
and
\begin{widetext}
\begin{eqnarray}
    &&G_4(x,y,z,w)=-6\lambda\int dx_1 G_2(x-x_1)G_2(x_1-y)G_2(x_1-z)G_2(x_1-w) \\ \nonumber
    &&-6\lambda\int dx_1G_2(x-x_1)\left[G_1(x_1)G_2(x_1-y)G_3(x_1-z,x_1-w)\right. \\ \nonumber
    &&\left.+G_1(x_1)G_2(x_1-z)G_3(x_1-y,x_1-w)
    +G_1(x_1)G_2(x_1-w)G_3(x_1-y,x_1-z)\right].
\end{eqnarray}
\end{widetext}
Now, we can check if our choice for $G_3(x,x,x)=0$ is correct. Indeed, we get
\be   
G_3(x,x,x)=-6\lambda\int dx_1 G_2(x-x_1)G_1(x_1)G_2(x_1-x)G_2(x_1-x)
\ee
where the product $G_2(x-x_1)G_2(x_1-x)$ appears with the Green function having opposite space-time support (they are different from zero in complementary manifolds and never overlap). On the integration volume, this yields 0, provided we ask for a causal behavior. This argument holds also for the other possible choices, as can be straightforwardly verified, giving a complete mathematical proof for our solution. 

Our situation is a favorable one as we have exactly solvable partial differential equations in our Dyson-Schwinger set. This can be extended without difficulty to a massive or a Higgs-like model \cite{Frasca:2015wva}. In our studies, we tried to extend the potential to a more general one $a\phi^2+b\phi^3+c\phi^4$ that was numerically analyzed in \cite{Adams:1993zs} with the aim to have all the constants known in the most general case. So far, we were not able to get a solution in closed form for it and this case remains only numerically settled from a quantum field theory standpoint.
%

\medskip

\section{Formalism}


In this section we present a set of formulas that will be of reference for our further computations. Most of them are can be found in textbooks \cite{Nair:2005iw,Peskin:1995ev} but we would like to provide a self-contained paper.

So, our aim is to obtain the effective potential as discussed in Ref. \cite{Coleman:1973jx} by recurring to the exact solution for the scalar field theory recently obtained in Ref.\cite{Frasca:2015yva}. This will allow us to understand the behavior of the effective potential also in the strong coupling limit.
Then, using the results presented in Ref.\cite{Frasca:2015yva}, we assume to have the partition function of a scalar field theory given by
\be
Z[j]=\sum_{n=1}^\infty\left(\prod_{m=1}^n\int d^4x_m\right) G_n(x_1..x_n)
\prod_{p=1}^nj(x_p),
\ee
with $G_n$ being the nP-functions, and one is able to get, in principle, all the nP-functions exactly. We can easily write down the effective action by noting that
\be
\phi_c(x)=\langle\phi(x)\rangle=\frac{\delta}{\delta j(x)}W[j]
\ee
with 
\be
W[j]=\ln Z[j].
\ee
Now, let us introduce the effective action by a Legendre transform
\be
\Gamma[\phi_c]=W[j]-\int d^4xj(x)\phi_c(x).
\ee
This is basically nothing but
\be
\Gamma[\phi_c]=\sum_{n=1}^\infty\left(\prod_{m=1}^n\int d^4x_m\right) \Gamma_n(x_1..x_n)
\prod_{p=1}^n\phi_c(x_p).
\ee
The first few relations between the nP-functions and the nP-vertices are given by
\begin{widetext}
\bea
\label{eq:Gn}
G_2(x_1,x_2)&=&\Gamma_2^{-1}(x_1,x_2), \nonumber \\
G_3(x_1,x_2,x_3)&=&\int d^4x'_1d^4x'_2d^4x'_3\Gamma_3(x'_1,x'_2,x'_3)G_2(x_1,x'_1)G_2(x_2,x'_2)G_2(x'_3,x_3),
\eea
\end{widetext}
and so on. It should be noted that
\begin{widetext}
\be
\int dz\frac{\delta^2\Gamma[\phi_c]}{\delta\phi_c(x)\delta\phi_c(z)}
\left(\frac{\delta^2\Gamma[\phi_c]}{\delta\phi_c(z)\delta\phi_c(y)}\right)^{-1}
=\delta^4(x-y).
\ee
\end{widetext}
Finally, we can evaluate the effective potential as  \cite{Nair:2005iw}
\begin{widetext}
\be
L^4V_{eff}[\phi_c]=-\sum_n\frac{1}{n!}\int d^4x_1\ldots d^4x_n\Gamma_n(x_1,\ldots,x_n)
(\phi_c(x_1)-\phi_0(x_1))\ldots(\phi_c(x_n)-\phi_0(x_n)).
\ee
\end{widetext}
It should be remembered that
\be
\left.\phi_c(x)\right|_{j=0}=\phi_0(x).
\ee
This corresponds to the 1P-function that in our case is not trivial while $\phi_c$ is taken to be constant. For a $\phi^4$ theory one starts from $\phi_0=0$. Finally, we give the same equation in momentum space as
\be
\label{eq:Vq}
V(\phi_c)=-\sum_n\frac{1}{n!}\left.\Gamma_n(q_1,\ldots,q_n)\right|_{q_i=0}(\phi_c-\phi_0)\ldots(\phi_c-\phi_0).
\ee

\medskip

\section{Standard Formalism}


The computation of bounce action involves a stationary phase approximation to write the false vacuum decay rate in order to express the solutions of the equations of motion in Euclidean space.
For the purpose of demonstration, let us take a real scalar field theory:
\begin{equation}
    S[\phi] = \int d^4 x \left[\frac{1}{2} (\partial \phi(x))^2 + V(\phi(x))\right],
\end{equation} 
Fate of this semi-classical vacuum depends on its decay rate ($\gamma$) per unit volume and can be related to the imaginary part of the false vacuum energy, which can be expressed in terms of path integral such as
\begin{align}
    \gamma = - 2\, \Im \,\mathcal{E}, \quad e^{- \mathcal{V} \mathcal{E}} \simeq Z = \int_{\phi_{FV}}^{\phi_{FV}} D \phi\, e^{- S[\phi]},
\end{align}
where $\mathcal{V}$ is the space-time volume, $\phi_{FV}$ denotes the false vacuum state, and the path integral is performed over all field configurations such that $\phi(x,\pm\infty) = \phi_{FV}$. This integral is basically expanded as a sum over saddle points (or stationary points), each of which correspond to solutions of the equations of motion, following Ref. \cite{Coleman:1977py,Callan:1977pt}.

As it is evident this sort of bounce formalism does not take into account radiative corrections, and, hence, is valid only for the tree-level action. However, when radiative corrections to the tree-level potential become important, 
there exist regions of \textit{bounce-like} field configurations which are not stationary points of the tree-level action but for which the phase varies sufficiently slowly that they nevertheless dominate the integral 
\cite{Iliopoulos:1974ur}
.

%
Now, as usual, if one incorporates radiative corrections and utilise effective potential instead, this is only applicable to weakly coupled theories as the latter is calculated in perturbation theory, where we aspect to see also corrections to the kinetic term. Moreover, there is no relation between the bounce formalism and the exact, non-perturbative effective action.

\medskip


\section{Non-perturbative Method for False Vacuum Decay}


Given the form of the effective action, we can evaluate the decay rate of the false vacuum of a scalar theory following Refs. \cite{Coleman:1977py,Callan:1977pt}. The decay rate is given by
\be
\gamma\sim L^{-4}e^{-\Gamma[\phi_b]}
\ee
where $L^{4}$ is the volume, $\Gamma$ the effective action,
and $\phi_b$ is obtained by
\be
\frac{\delta\Gamma}{\delta\phi_c}=0.
\ee
In order to evaluate it, we need to compute the effective potential given in eq.(\ref{eq:Vq}). 
We do this by exploiting the exact solutions for the 1P- and 2P-functions.
Following the techniques developed in Ref. \cite{Frasca:2015yva} (for a brief review of the technique, see supplementary materials),
\begin{widetext}
\bea
G_3(x-x_1,x-x_2)=-\int d^4x_3G_2(x-x_3)V'''[\phi_0(x_3)]G_2(x_1-x_3)G_2(x_2-x_3)=&& \nonumber \\
-\int d^4x_1d^4x_2d^4x_3G_2(x-x_3)V'''[\phi_0(x_3)]\delta^4(x-x_1)\delta^4(x-x_2)G_2(x-x_3)G_2(x-x_3).&&
\eea
This equation yields immediately $\Gamma_3$ by comparing its the last line directly with
eq.(\ref{eq:Gn}), providing
\be
\Gamma_3(x_1,x_2,x_3)=-V'''[\phi_0(x_1)]\delta^4(x_1-x_2)\delta^4(x_1-x_3).
\ee
\end{widetext}
This implies, using eq.(\ref{eq:Vq}) and stopping at the third term, that the full potential has the form
\be
\label{eq:FullU}
U(\phi_c)=V(\phi_c)+\frac{1}{2}V_2(\phi_c-\phi_0)^2-\frac{1}{3!}|V_3|(\phi_c-\phi_0)^3.
\ee

Let us evaluate the effective potential for the $\phi^4$ theory. We will have for the 2P-vertex function in momenta space
\be
\Gamma_2(p)=\left(\sum_{n=0}^\infty\frac{B_n}{p^2-m_n^2+i\epsilon}\right)^{-1},
\ee
where
\be
B_n=\frac{\pi^3}{4K^3(i)}(2n+1)^2\frac{e^{-\left(n+\frac{1}{2}\right)\pi}}{1+e^{-(2n+1)\pi}},
\ee
with $K(i)$ being the complete elliptic integral of the first kind, 
and
\be
m_n=(2n+1)\frac{\pi}{2K(i)}(\lambda/2)^\frac{1}{4}\mu,
\ee
being the mas spectrum. This will yield immediately
\be
V_2=\frac{K(i)}{2\pi}\mathscr{A}\sqrt{2\lambda}\mu^2=\sqrt{2\lambda}\mu^2,
\ee
with\footnote{Another interesting identity is given by $\left(\sum_{n=0}^\infty (2n+1)^2A_n\right)^{-1}=4K^3(i)/\pi^3$.}
\be
\mathscr{A}=\left(\sum_{n=0}^\infty A_n\right)^{-1}=\frac{2\pi}{K(i)},
\ee
and
\be
A_n=\frac{e^{-\left(n+\frac{1}{2}\right)\pi}}{1+e^{-(2n+1)\pi}}.
\ee
Then, one has
\be
V_3=-6\lambda\phi_0(0)=-3(2\lambda)^\frac{3}{4}\mu\operatorname{sn}(\theta,i).
\ee
Here, $\theta$ is an integration constant of the equation for the 1P-correlation function.

Now, we can write the Euclidean effective action as
\be
\Gamma_E(\phi_c)=-\int d^4x\left[\frac{1}{2}(\partial\phi_c)^2+U(\phi_c)\right]
\ee
with the effective potential looking like
\be
\label{eq:Ueff}
U(\phi_c)=
\frac{1}{2}V_2(\phi_c-\phi_0)^2+\frac{1}{3!}V_3(\phi_c-\phi_0)^3+\frac{\lambda}{4}\phi_c^4.
\ee
In this formula, we can neglect the $\phi_0$ contribution in the formal limit $\lambda\rightarrow\infty$ as it goes like $\lambda^{-\frac{1}{4}}$. For the extrema, of the potential 
we get the following values in the strong coupling limit,
\be
2^\frac{3}{4}k_1\mu\lambda^{-\frac{1}{4}}, \quad 0, \quad -2^\frac{3}{4}k_2\mu\lambda^{-\frac{1}{4}},
\ee
where $k_1=k_1(\theta)$ and $k_2=k_2(\theta)$ are numerical constants depending on the initial phase $\theta$ given by
\bea
k_1(\theta)&=&3\operatorname{sn}(\theta,i) + \sqrt{9\operatorname{sn}^2(\theta,i) - 8}, \nonumber \\
k_2(\theta)&=&
-3\operatorname{sn}(\theta,i) + \sqrt{9\operatorname{sn}^2(\theta,i) - 8}.
\eea
We can now see that the only condition for the existence of the extrema is
\be
\label{eq:lim}
|\operatorname{sn}(\theta,i)|>\sqrt{\frac{8}{9}}\approx 0.94.
\ee
This means that the phase should be chosen to have $|\operatorname{sn}(\theta,i)|$ near 1. The physical implication is that the symmetry $\phi\rightarrow -\phi$ is spontaneously broken.

We observe in this computation the non-trivial fact that we get a cubic term with the right sign, implying the breaking of the $Z_2$ symmetry. On the other side, this comes out as no surprise as already at the level of the Dyson-Schwinger set of equations we see the appearance of a mass term originating from quantum fluctuations (see eq.(\ref{eq:dm})) that breaks the conformal invariance of the theory. Such results reflects a very general principle that in Nature whichever symmetry is present it tends to get broken by quantum effects unless some screening mechanism is present.
%

Our aim is not to exploit the formalism in the most general case but just to see if, in some manageable approximation, an improved understanding of known results could be achieved for the false vacuum decay. Therefore, we apply the thin-wall approximation
Ref.\cite{Coleman:1977py,Weinberg:1996kr}. One has
\be
\gamma\propto e^{-B}
\ee
where
\be
B=\frac{27\pi^2\mathscr{S}^4}{2\epsilon^3}
\ee
with
\be
\mathscr{S}=\int_0^{\langle\phi_c\rangle}\sqrt{2U(f)}df
\ee
and
\be
\epsilon = U(\langle\phi_c\rangle).
\ee
We will have $\epsilon = c_1\mu^4$ with $c_1$ being a numerical constant. This allows us to evaluate, for $\lambda$ being large enough, by the trapezoidal rule improved by estimating at an extremum\footnote{The leading error is proportional to $|f'(b)-f'(a)|$, being $f(x)$ the function to integrate and $b$ and $a$ the extremes of integration. In our case is $f'(a)=0$ and $f'(b)=0$ being $b$ an extremum of the potential.},
\be
\mathscr{S}\approx\frac{1}{2}\sqrt{2U(\langle\phi_c\rangle)}\langle\phi_c\rangle=\frac{1}{2}\sqrt{2\epsilon}\langle\phi_c\rangle.
\ee
Assuming $\rm{sn}(\theta,i)\approx 1$, we get
\be
\epsilon=\kappa_2(\theta)\frac{\mu^4}{16},
\ee
where
\begin{widetext}
\be
\kappa_2(\theta)=
-27\operatorname{sn}^4(\theta,i) + 9\sqrt{9\operatorname{sn}^2(\theta,i) - 8}\operatorname{sn}^3(\theta,i) + 36\operatorname{sn}^2(\theta,i) - 8\sqrt{9\operatorname{sn}^2(\theta,i) - 8}\operatorname{sn}(\theta,i) - 8.
\ee
\end{widetext}
Then,
\be
\label{eq:B1}
B=\frac{432\pi^2}{\lambda}\frac{k_2^4(\theta)}{\kappa_2(\theta)}
\ee
We have assumed $\lambda>0$ everywhere. It is important to emphasize that, in the interesting range for $\theta$, both $k_2(\theta)$ and $\kappa_2(\theta)$ are always greater than 1. Similar behaviour can be observed for $k_1(\theta)$ and $\kappa_1(\theta)$. For $\operatorname{sn}(\theta,i)=1$, we get the formula
\be
\label{eq:B2}
B=3456\frac{\pi^2}{\lambda},
\ee
that holds for $\lambda\gg 1$. The final decay rate becomes
\be
\gamma\sim L^{-4}e^{-\frac{432\pi^2}{\lambda}\frac{k_2^4(\theta)}{\kappa_2(\theta)}}
\approx L^{-4}e^{-3456\frac{\pi^2}{\lambda}},
\ee
and with the approximation $\operatorname{sn}(\theta,i)=1$ that satisfies the aforementioned condition $|\operatorname{sn}(\theta,i)|>\sqrt{8/9}\approx 0.94$.

\medskip

\section{Results}


The relevant result of our analysis is the appearance of minima lower than the expected ground state of the theory we started with. Indeed, this theory is conformally invariant at the classical level. Quantum fluctuations destroy this symmetry. The following plots 
yield a visual representation
of our results. We emphasize that we are using the thin wall approximation, firstly devised by Coleman in \cite{Coleman:1977py}, for comparison reasons with the textbook approach. This approximation was discussed recently in \cite{Brown:2017cca}.  To understand its relevance in our case, let us consider the potential written in the form
\be
U(\phi_c)=\frac{\lambda}{4}\phi^4_c+U_B(\phi_c),
\ee
where $U_B(\phi_c)$ are the terms that break the symmetry. From eq.(\ref{eq:Ueff}), one gets
\be
U_B(\phi_c)=-\frac{1}{2}(2\lambda)^{3/4}\mu\operatorname{sn}(\theta,i)\phi_c^2\left(\phi_c-\frac{\mu}{(2\lambda)^{1/4}\operatorname{sn}(\theta,i)}\right)
\ee
and this yields the difference of energy-density between true and false vacuum as
\be
\label{eq:eps}
\epsilon=2\lambda\mu^2\operatorname{sn}^2(\theta,i)\phi_c^2.
\ee
This provides the range of validity of our approximation.
%

\begin{figure}[H]
\centering
\includegraphics[height=5cm,width=5cm]{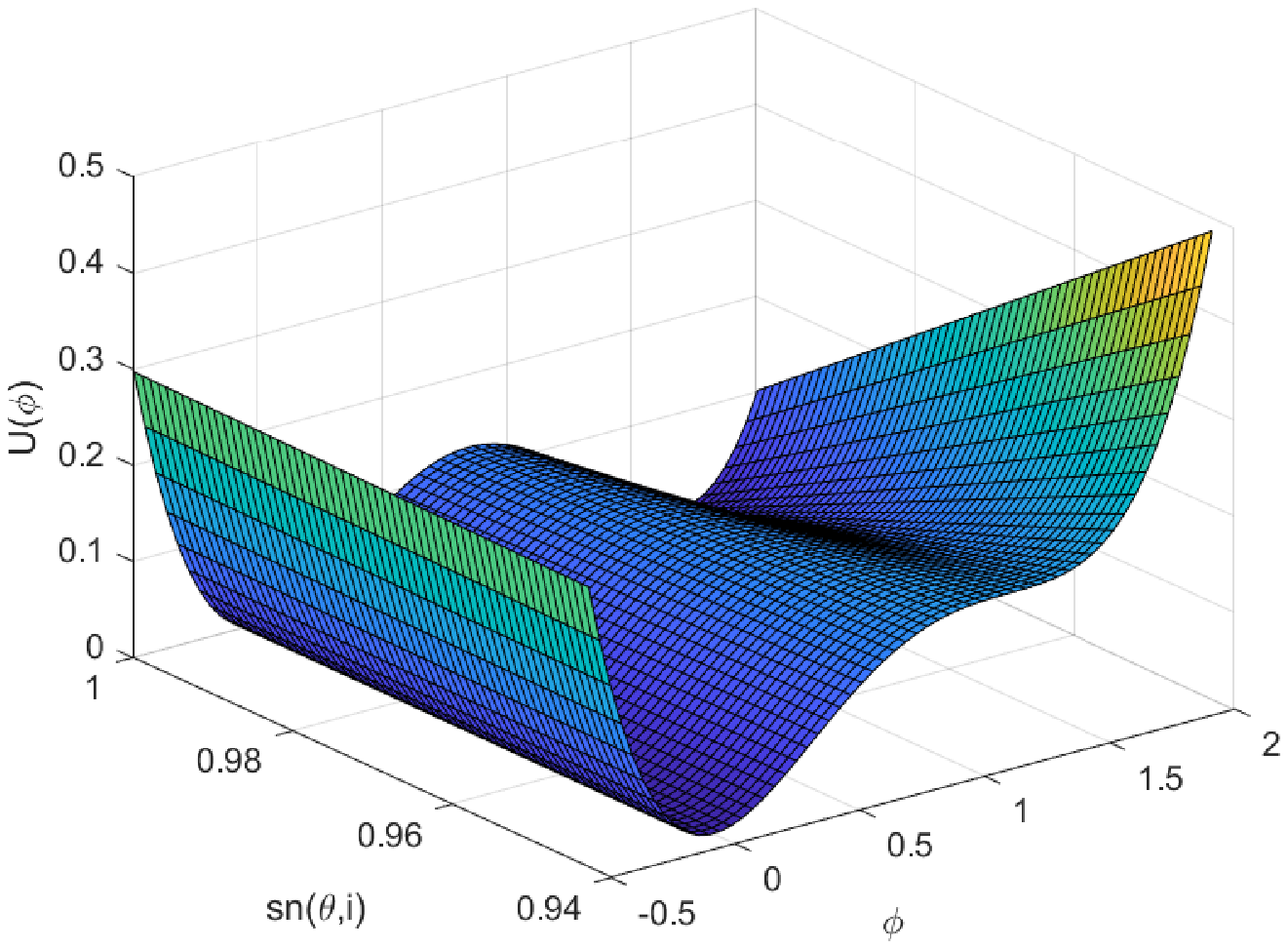}
\caption{\it Plot of $U(\phi)$ assuming a fixed $\lambda=1$ and varying the phase in the permitted range given in eq.(\ref{eq:lim}). We have taken $\mu=1\ \rm{TeV}$ just to fix the energy scale.\label{fig1}}
\end{figure}

\begin{figure}[H]
\centering
\includegraphics[height=5cm,width=5cm]{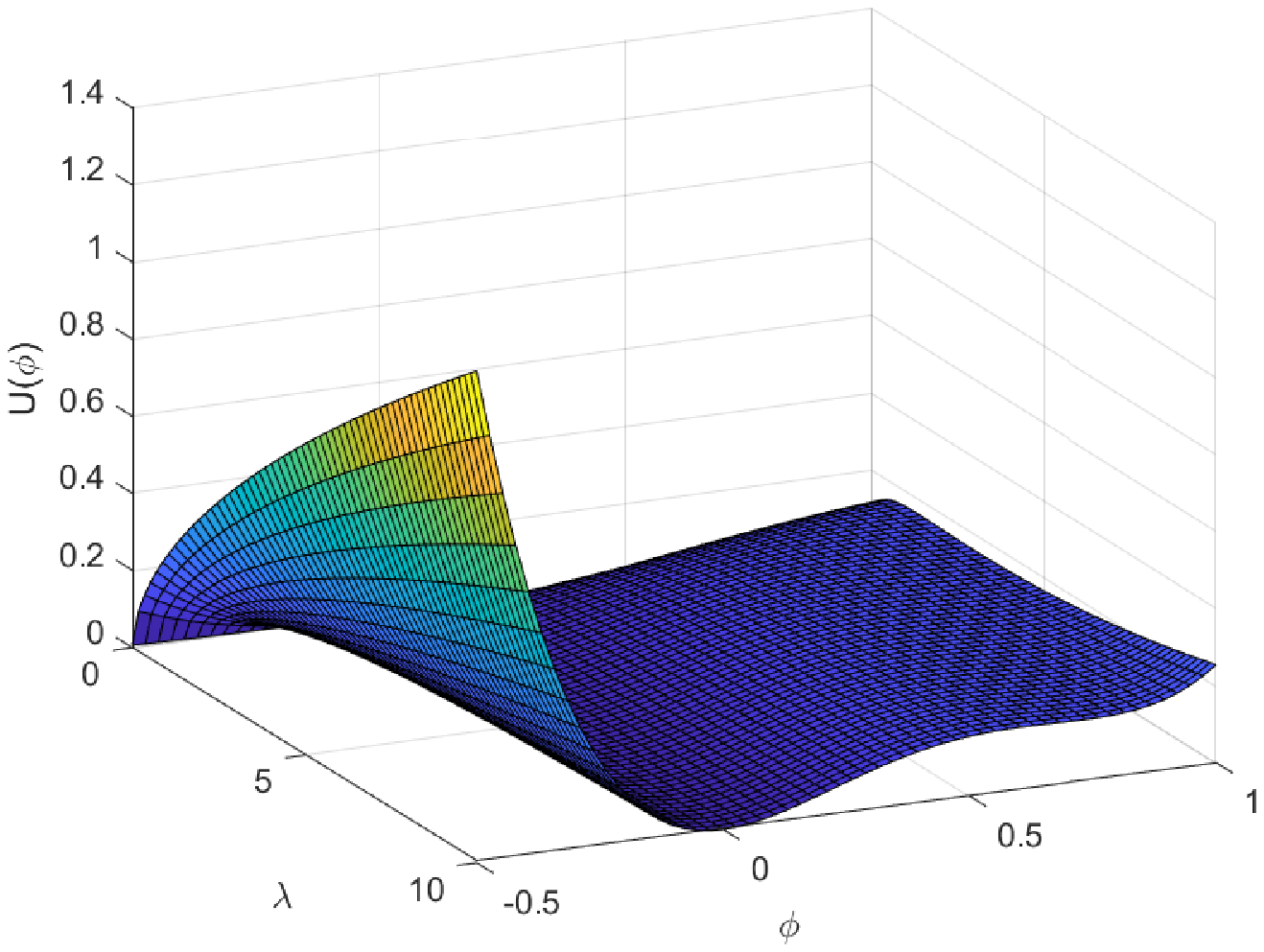}
\caption{\it The potential $U(\phi)$ assuming a fixed $\theta$ and varying the coupling. Also here, we have taken $\mu=1\ \rm{TeV}$.\label{fig2}}
\end{figure}

\begin{figure}[H]
\centering
\includegraphics[height=6cm,width=8cm]{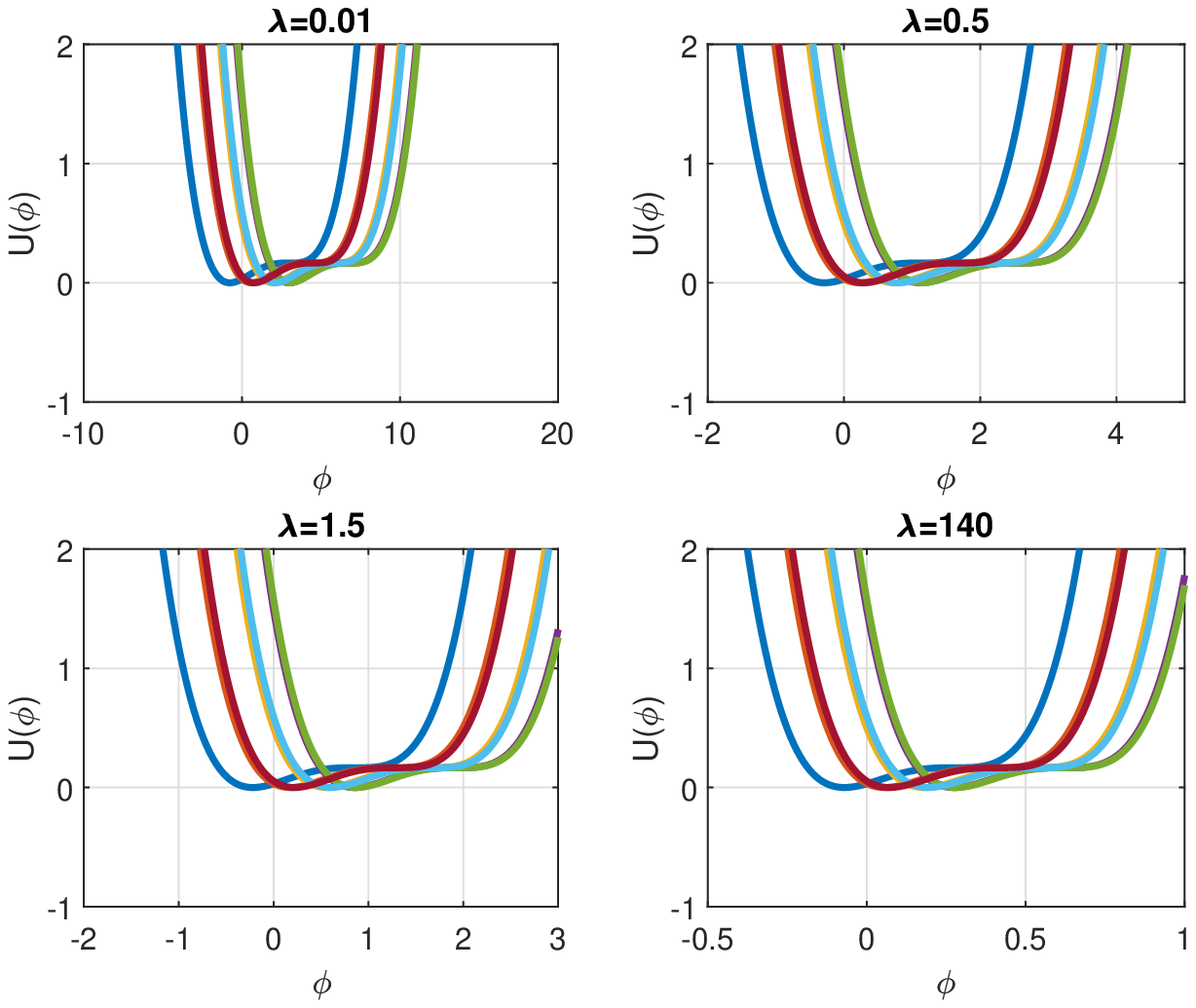}
\caption{\it Plot of $U(\phi)$ as given in eq.~(\ref{eq:FullU}) given $\phi_0$ being the expected vacuum solution. It is seen that the behavior is kept notwithstanding the varying background $\phi_0$ at varying coupling $\lambda$. Different colors represent different curves plotted at various values given by the background $\phi_0$ (see eq.(\ref{solG1}) in the Supplementary Materials) so to give the corresponding spreading. We note that such a spreading decreases with increasing $\lambda$ showing the effect of non-perturbative regimes. 
\label{fig3} 
}
\end{figure}

Varying the phase $\theta$, we have Fig.~\ref{fig1}.
\noindent Similarly, fixing the phase $\theta$ to a value in the permitted range, we get Fig.~\ref{fig2}.
\noindent We can have an understanding of the full potential, so that we can consider the full range of $\lambda$, from Fig.\ref{fig3}.
\noindent From Fig.\ref{fig3}, it is clear that 
the behavior becomes more and more stable as the coupling $\lambda$ increases with a space-time varying background. This can be also understood through our derivation of eq.(\ref{eq:eps}) where one can see that more and more energy is needed to allow the vacuum decay.



\section{Conclusions}

\textit{In a nut-shell}, we developed a new approach to compute the effective potential for a scalar field theory. We used exact background solutions that are known for the $\phi^4$ theory and showed how a $\phi^3$ term with negative sign may arise naturally from our analysis, leading to a false vacuum and its decay. In this manner, even if we started from a scale-invariant $\lambda \phi^4$ theory, we are left with two minima and a meaningful computation for the false vacuum decay in the strong coupling regime. All the coefficients can be properly fixed and depends solely on the quartic coupling with various functional dependencies. 
Our main findings are:
\begin{itemize}
    \item Since the effective potential we obtained was based on exact solutions and exact Green's functions, our results hold for both small and large coupling values (as shown in eq.(\ref{eq:Ueff}) and in Fig.\ref{fig3} ).
    \item One may imagine that the Higgs mass here is dynamically generated due to strong coupling dynamics and the ambiguity for the resulting Higgs mass is from the integration constants that can be taken to have the right values. We have used $\mu = 1$ TeV here.
    \item We showed that in the strong coupling regime, the quartic interaction term dynamically generates a cubic term and a quadratic term, leading to creating and decay of the false vacuum.
    \item False vacuum decay rate was evaluated by following our exact Green's function in the non-perturbative regime, as given in eqs.(\ref{eq:B1}) \& (\ref{eq:B2}). 
\end{itemize}
    We established, following our exact Green's functions, that quantum field theory is no more limited to small perturbation theory that limits its possible range of applications and can be nicely extended to non-perturbatve strongly-coupled BSM theories, following the effective potential approach. In particular, we obtained results for the fate of false vacuum in the non-perturbative regimes and envisage this will enable us to find numerous applications in primordial cosmology like inflation and for first-order phase transition with predictable signals in Gravitational Waves, searched for, and with great detection prospects in LIGO and VIRGO \cite{Romero:2021kby,LIGO-result}. The collaboration has already been able to put strong limits on stochastic Gravitational Waves background (SGWB) from such primordial sources. However, a detailed computation of phase transition and Gravitational Waves following our non-perturbative method is beyond the scope of the present analysis and we plan to take it up soon. Finally we envisage the methodology we developed will also be applicable beyond the high energy physics community to condensed matter systems particularly that of ultracold atoms and Bose gas \cite{Billam:2021nbc} and quantum spin chain systems \cite{Lagnese:2021grb}. 

\section*{Acknowledgement}

We would like to thank Marek Lewicki, Alberto Salvio and Alessandro Strumia for very useful comments.
This work is supported in part by the United States, Department of Energy Grant No. DE-SC0012447 (N.O.).

\medskip

\section*{Appendix}

We show that the breaking of translation invariance of the theory entails a zero mode. The Hamiltonian of the system is given by
\begin{equation}
    H=\int d^3x\left[\frac{1}{2}(\partial_t\phi)^2+\frac{1}{2}(\nabla\phi)^2+\frac{\lambda}{4}\phi^4\right].
\end{equation} 
By linearizing around the classical solution (given by $G_1$ in (\ref{solG1}))
\begin{equation}
    \phi(x)=\phi_0(x)+\delta\phi(x)\,,
\end{equation}
yielding
\begin{equation}
   H=H_0+\int d^3x\left[\frac{1}{2}(\partial_t\delta\phi)^2+\frac{1}{2}(\nabla\delta\phi)^2
   +\frac{3}{2}\lambda\phi_0^2\delta\phi^2\right]+O\left(\delta\phi^3\right),
\end{equation}
being $H_0$ the contribution coming from the classical solution. The linear part can be diagonalized with a Fourier series provided we are able to get the eigenvalues and the eigenvectors of the operator
\begin{equation}
   L_{\mu_0^2=0}=-\Box+3\lambda\phi_0^2(x).
\end{equation}
It is not difficult to realize that there is a zero mode. We give the solutions for both the zero and non-zero modes. The spectrum is continuous with eigenvalues 0 and $3\mu^2\sqrt{\lambda/2}$ with $\mu$ varying continuously from 0 to infinity. The zero-mode solution has the aspect
\begin{equation}
  \chi_0(x,\mu)=a_0\,{\rm cn}(p\cdot x+\theta,i)\,{\rm dn}(p\cdot x+\theta,i)
\end{equation}
being $a_0$ a normalization constant. Non-zero modes are given by
\begin{equation}
  \chi(x,\mu)=a'\,{\rm sn}(p\cdot x+\theta,i)\,{\rm dn}(p\cdot x+\theta,i).
\end{equation}
with $a'$ again a normalization constant. These hold on-shell, that is when $p^2=\mu^2\sqrt{\lambda/2}$. Since the spectrum is continuous, these eigenfunctions are not normalizable. Therefore, we note that there is a doubly degenerate set of zero modes spontaneously breaking translational invariance and the $Z_2$ symmetry of the theory. 
This gives for the zero mode
\begin{equation}
  \chi_0(x,\mu)=-2a_0\frac{{\rm sn}(p\cdot x,i)}{{\rm dn}^2(p\cdot x,i)}.
\end{equation}

\end{document}